\documentclass[a4paper,12pt]{article}
\usepackage{multicol}
\usepackage[left=1.5cm,right=1.5cm,top=2cm,bottom=2cm]{geometry}

\usepackage{ccaption} 
\captiondelim{. } 
\usepackage{cite}

\usepackage{amsfonts,amssymb,amsthm,mathtools} 
\usepackage{amsmath}
\usepackage{icomma} 
\usepackage{euscript}	 
\usepackage{mathrsfs} 
\usepackage{txfonts}

\usepackage{graphicx}  
\usepackage{subfig}
\usepackage{float}
\usepackage{wrapfig} 

\usepackage{array,tabularx,tabulary,booktabs} 
\usepackage{longtable}  
\usepackage{multirow} 

\title{THE GENERATION OF NARROW ULTRARELATIVISTIC BEAMS OF POSITRONS (ELECTRONS) IN THE PROCESS OF RESONANT PHOTOGENERATION OF PAIRS
	ON NUCLEI IN A STRONG ELECTROMAGNETIC FIELD}
\author{S. P. Roshchupkin$^1$, S. S. Starodub$^2$\\\\
$^1$ Peter the Great St. Petersburg Polytechnic University,\\
195251, St-Petersburg, Russian Federation, Russia\\
$^2$ Institute of Applied Physics, National Academy of Sciences of Ukraine,\\
 40000, Sumy, Ukraine} 
 
\date{}

\begin{document} 

\maketitle


\begin{abstract}
The generation of narrow beams of high-energy positrons (electrons) in the process of resonant photogeneration of ultrarelativistic electron-positron pairs by high-energy gamma quanta in the field of the nucleus and a strong electromagnetic wave is theoretically predicted. It is shown that if the energy of the initial gamma quanta significantly exceeds the characteristic energy of the process, then ultrarelativistic positrons (for channel A) or electrons (for channel B) are emitted with energies very close to the energy of gamma quanta. Moreover, the resonant differential cross-section of such processes can exceed the corresponding differential cross-section without an external field by thirteen orders of magnitude.  This effect makes it possible to obtain narrow beams of ultrarelativistic positrons (electrons) in strong electromagnetic fields with high probability.
\end{abstract}

\section{Introduction}
Currently, laser systems of high intensities are being intensively created \cite{1,2,3,4,5,6,7,8}, as well as sources of high-energy particles \cite{9,10,11,12}. This contributes to the intensive development of quantum electrodynamics (QED) in strong electromagnetic fields \cite{13,14,15,16,17,18,19,20,21,22,23,24,25,26,27,28,29,30,31,32,33,34,35,36,37,38,39,40,41,42,43,44,45,46,47,48,49,50,51,52,53,54,55}. An important place in such processes is occupied by resonant effects (Oleinik resonances  \cite{24,15}) associated with the release of an intermediate particle in an external electromagnetic field onto the mass shell (see, for example, articles \cite{26,27,28,29,30,31,32,33,34,35,36,37}.]). It is important to emphasize that the resonant differential cross-sections can significantly exceed the corresponding non-resonant differential cross-sections \cite{29,30,31,32,33,34,35,36,37,38}. In a recent paper  \cite{36}, the resonant photogeneration of electron-positron pairs on nuclei in strong light fields was studied. At the same time, however, the possibility of generating positrons (electrons) with energies close to the energies of the initial gamma quanta has not been studied.

We will study the resonant process of photogeneration of pairs (PGP) for high-energy initial gamma quanta, as well as produced electrons and positrons when the basic classical parameter
\begin{equation} \label{eq1}
	\eta =\frac{eF\lambdabar}{mc^2}
\end{equation}
satisfies the relation
\begin{equation} \label{eq2}
	\eta \ll \frac{\hbar \omega_i}{mc^2}\gg 1, \qquad \eta \ll \frac{E_{\pm}}{mc^2}\gg 1.
\end{equation}
Here $e$ and $m$ are the charge and mass of the electron,  and   $F$ and\quad$\lambdabar=c/\omega$  are the electric field strength and wavelength, $\omega $ is the frequency of the wave;  $\omega_i$  is the high-energy initial gamma quantum, $E_{\pm}$  is the ultrarelativistic energy of the positron (electron).

The article \cite{36} shows that the resonant energy of the produced positron (for channel A) and electron (for channel B) is determined by their outgoing angles (see Exps. (\ref{eq10}, \ref{eq12})), as well as the quantum parameter 
\begin{equation} \label{eq3}
\varepsilon_{\eta\left(r\right)} =r \varepsilon_\eta \ge 1, \quad \varepsilon_\eta=\frac{\omega_i}{\omega_\eta},	
\end{equation}
Here the parameter $\varepsilon_{\eta\left(r\right)}$  is numerically equal to the product of the number of absorbed photons in the external field stimulated Braith-Wheeler process $\left(r=1,2,3 \dots\right)$  by the quantum parameter $\varepsilon_\eta$. This parameter is equal to the ratio of the energy of the high-energy initial gamma quantum to the characteristic energy of the process $\hbar \omega_\eta$  which is determined by the experimental conditions and the laser installation: 
\begin{equation} \label{eq4}
	\hbar \omega_\eta =\frac{\left(mc^2\right)^2\left(1+\eta^2\right)}{\left(\hbar \omega\right)\sin^2\left(\theta_i/2\right)}.
\end{equation}
Here  $\theta_i$ is the angle between the momentum of the initial gamma quantum and the direction of wave propagation. It can be seen from expression (\ref{eq4}) that the value of the characteristic energy $\hbar \omega_\eta$  is inversely proportional to the photon energy of the wave $\left(\hbar \omega\right)$  and is also directly proportional to the intensity of the wave $\left(I\sim \eta^2\ \left(\mbox{Wcm}^{-2}\right)\right)$. It is important to note that the number of absorbed photons under resonance conditions significantly depends on the range of values of the quantum parameter  $\varepsilon_\eta$ (\ref{eq3}) or the relationship between the characteristic energy of the process and the initial energy of the gamma quantum. If  $\varepsilon_\eta < 1 \left(\omega_i < \omega_\eta\right)$, then the number of absorbed photons of the wave in the external field-stimulated Braith-Wheeler process must exceed or be equal to some minimum number of photons of the wave $r_{\min}$, which is determined by the parameter $\varepsilon_\eta$ (see the equation (\ref{eq13})). At the same time, the number of absorbed photons of the wave can start with quite large numbers when $r_{\min} \gg 1 \left(\omega_i \ll \omega_\eta\right)$ (this is usually true for very strong electromagnetic fields). If it is a quantum parameter $\varepsilon_\eta \ge 1 \left(\omega_i \ge \omega_\eta\right)$, then the number of absorbed photons of the wave always begins with one photon (see the equation (\ref{eq14})). It is important to emphasize that the number of absorbed photons of the wave significantly affects the magnitude of the resonant differential cross section. For a small number of absorbed photons of the wave $\left(r \sim 1\right)$ , the resonant cross section will be significantly larger than for a large number of absorbed photons  $\left(r \gg 1\right)$. Because of this, the case when the energy of the initial gamma quanta exceeds the characteristic energy of the process is of undoubted interest. Note that case (\ref{eq13})  was studied in detail in \cite{36}. At the same time, case (\ref{eq14}) was not considered.

In this paper, within the framework of the ratio (\ref{eq14}) we will mainly study consider the case
\begin{equation} \label{eq5}
\omega_i \gg \omega_\eta \quad \left(\varepsilon_{\eta\left(r\right)}=r\varepsilon_\eta \gg 1, r=1,2,3 \dots \right),
\end{equation}
when the resonant generation of ultrarelativistic positrons (electrons) takes place with maximum probability and with energies close to the energies of the initial gamma quanta.

We will use the relativistic system of units: $\hbar=c=1$.

\section{Resonant energies of positrons (electrons) in strong fields}

Oleinik resonances occur when an intermediate electron (positron) in the electromagnetic wave field enters the mass shell \cite{24,25,36}. Because of this, for channels A and B, we get (see Figure~\ref{figure1}):
\begin{figure}[H]
	\centering
	\includegraphics[width=7cm]{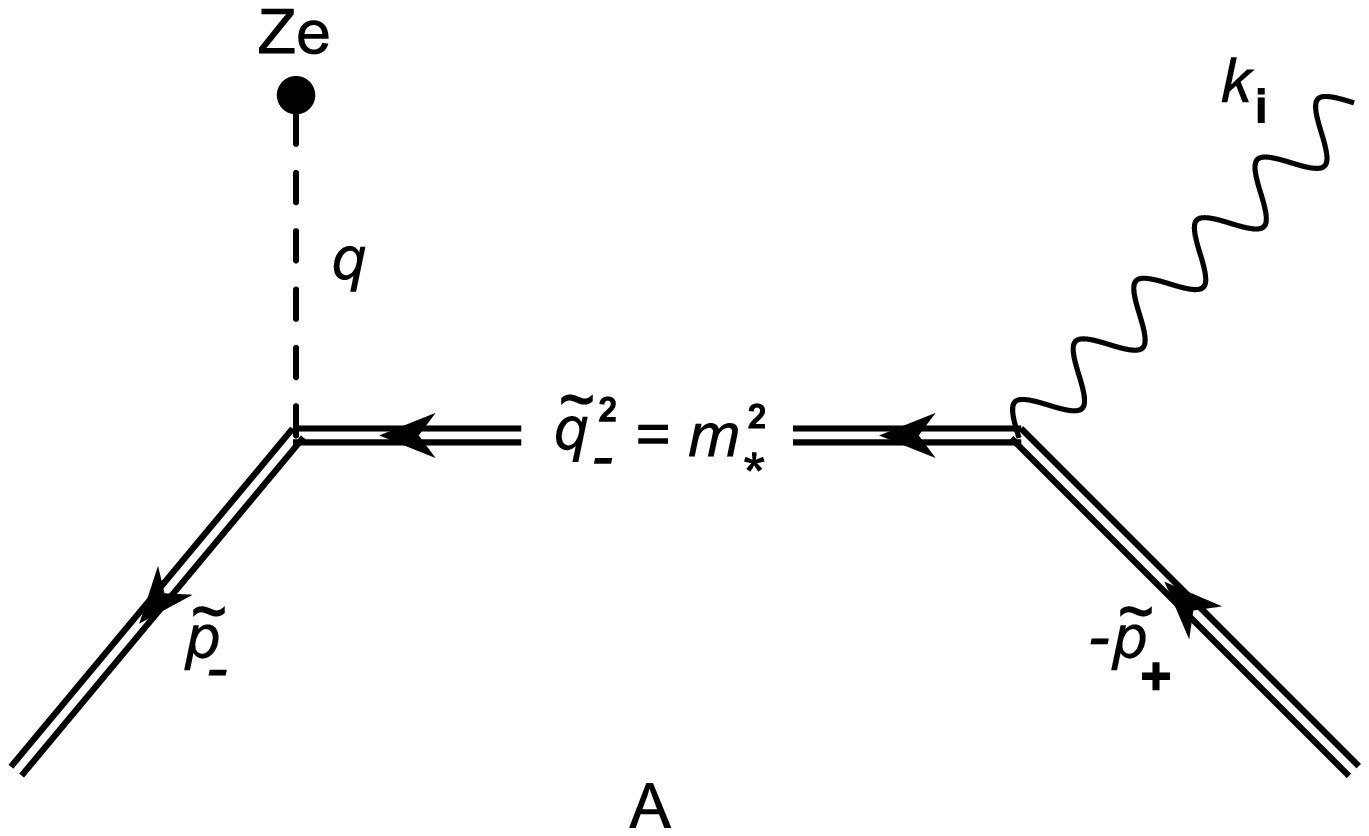} \qquad 
	\includegraphics[width=7cm]{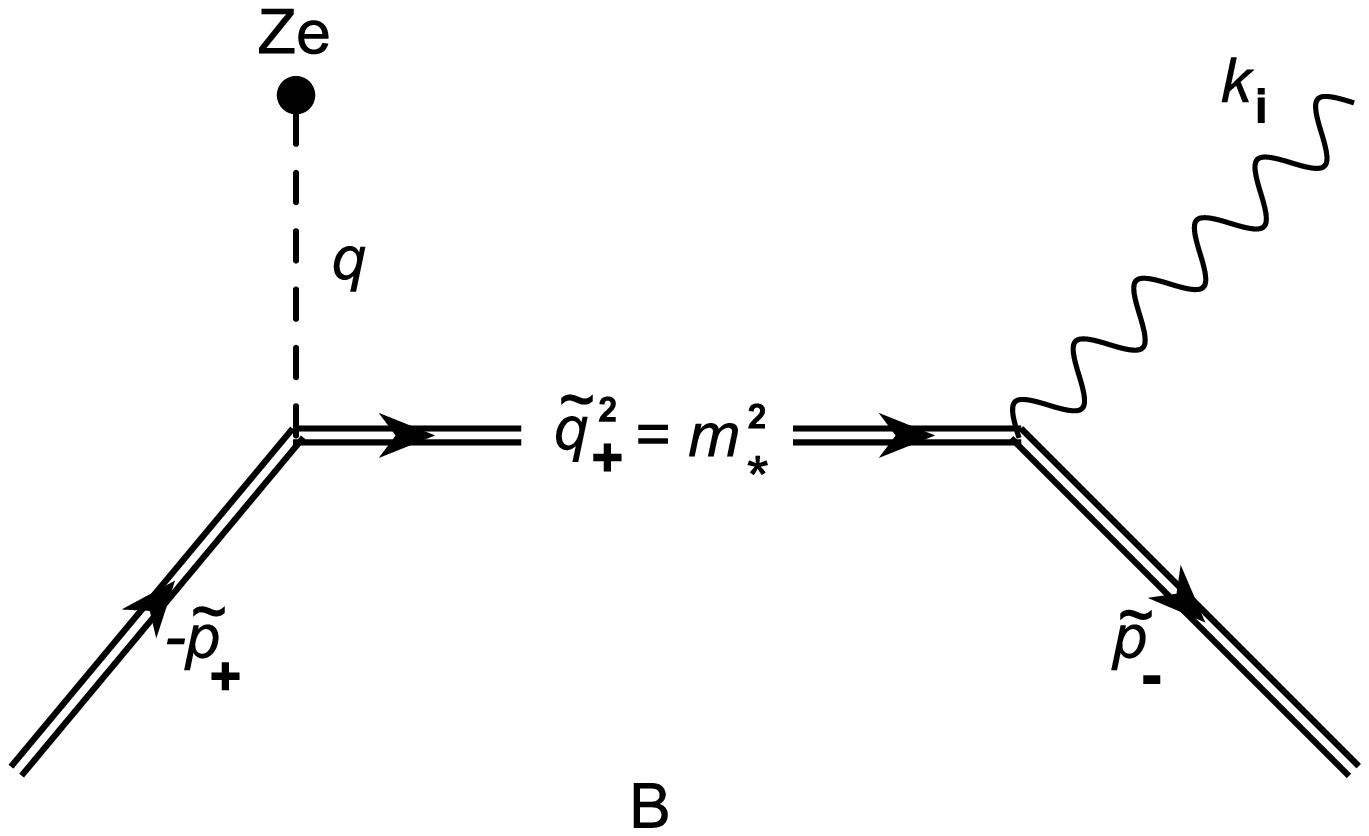}
	\caption{Resonant photogeneration of electron-positron pairs on nuclei in the field of a plane electromagnetic wave.}
	\label{figure1}
\end{figure}  
\unskip
\begin{equation}\label{eq6}
	\tilde q^2_{\mp}=m^2_*,\ \tilde q_{\mp}=k_i-\tilde p_{\pm}+rk.
\end{equation}
Here $\tilde q_-$  and $\tilde q_+$ are the 4-quasimomenta of intermediate electron (for channel A) and intermediate positron (for channel B), $m_*$  is the effective mass of the electron (positron) in the field of a circularly polarized wave \cite{23,36}
\begin{equation}\label{eq7}
	\tilde p_{\pm}=p_{\pm}+\eta^2 \frac{m^2}{2\left( kp_{\pm} \right)}k,\quad \tilde q_{\mp}=q_{\mp}+\eta^2 \frac {m^2}{2\left(kq_{\mp} \right)}k,\quad j=i,f.
\end{equation}
\begin{equation}\label{eq8}
	\tilde p^2_{\pm}=m^2_*,\quad m_*=m\sqrt{1+\eta^2}.
\end{equation}
In expressions (\ref{eq6})-(\ref{eq7}) $k=\left(\omega, \mathbf{k} \right)$ is the 4-momentum of the external field photon, $ p_{\pm}=\left(E_{\pm},\mathbf{p}_{\pm}\right)$ is the 4-momentum of the positron (electron). Such behavior is caused by the quasi-discrete energy spectrum of fermion propagating within the plane electromagnetic wave. Due to that fact, one may interpretate it as the reduction of the second order process into the two successive second order processes in fine structure constant (see Figure~\ref{figure1}).

In this paper, we examine the case of high-energy energies of the initial gamma quantum (\ref{eq2}). Moreover, we confine ourselves with the configuration, where all produced ultrarelativistic particles propagate within the narrow cone with the initial gamma quantum direction. Additionally, we demand those directions of initial gamma quantum and external wave propagation do not coincidence, otherwise resonances are impossible \cite{29,30,36}:
\begin{eqnarray}\label{eq9}
	 \theta_{i\pm}  =\measuredangle \left( \mathbf{k_i},\mathbf p_{\pm} \right)\ll 1,\quad & \overline{\theta}_{\pm} =\measuredangle \left( \mathbf p_-,\mathbf p_+ \right) \ll 1, \nonumber \\	
	 \theta_i  =\measuredangle \left( \mathbf{k_i},\mathbf k \right)\sim 1,\quad & \theta_{\pm}=\measuredangle \left( \mathbf k,\mathbf p_{\pm} \right)\sim 1.	
\end{eqnarray}

In this paper, we will consider the energies of the initial gamma quantum $\omega_i \lesssim 10^3 \mbox{GeV}$ and also in a wide range of photon energies of an electromagnetic wave $\left(1 \mbox{eV} \lesssim \omega \lesssim 10^4 \mbox{eV}\right)$. At the same time, we will consider the intensities of the electromagnetic wave significantly less than the critical intensities of the Schwinger $\left(I\ll I_* \sim 10^{29} \mbox{Wcm}^{-2}\right)$.

\begin{figure}[H]
 	\centering
 		\includegraphics[width=10cm]{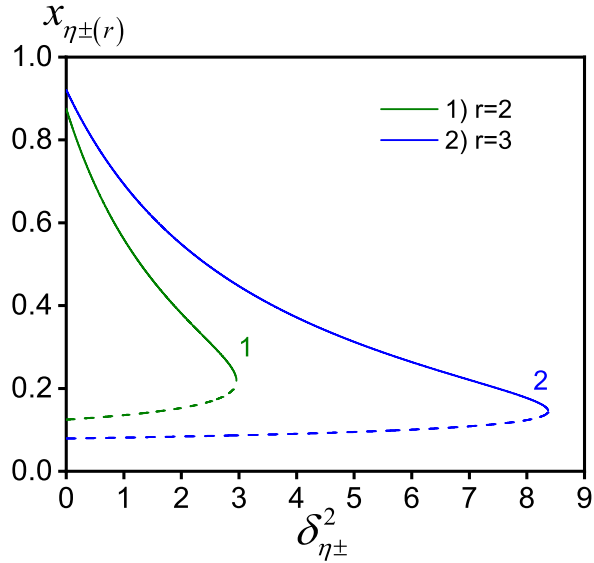}
 	\caption{Resonant positron (for channel A) and electron (for channel B) energies as functions of corresponding outgoing angles (at fixed $\omega_i = 60 \mbox{GeV}, \ \omega = 10 \mbox{eV}, \ \theta_i = \pi$). The solid lines represent high-energy solutions, the dashed lines stand for low-energy solutions. Curves 1 and 2 are plotted for $r=2, \ 3$, correspondingly.}
 	\label{figure2}
 \end{figure}

We determine the resonant energy of the positron $\left(E_{\eta + \left(r\right)}\right)$ (for channel A, see Figure~\ref{figure1}~A) and electron  $\left(E_{\eta - \left(r\right)}\right)$ (for channel B, see Figure~\ref{figure1}~B). We take into account the relations (\ref{eq9}) in the resonant condition (\ref{eq6}). After simple calculations, we get \cite{36}
\begin{equation}\label{eq10}
	x_{\eta j\left( r \right)}=\frac{\varepsilon_{\eta\left(r\right)}\pm \sqrt{\varepsilon_{\eta\left(r\right)}\left(\varepsilon_{\eta\left(r\right)}-1\right)-\delta^2_{\eta j}}}{2\left(\varepsilon_{\eta\left(r\right)}+\delta^2_{\eta j}\right)}, \quad j=\pm.
\end{equation}
Here it is indicated:

\begin{equation}\label{eq11}
x_{\eta \pm \left( r \right)}=\frac{E_{\eta \pm \left( r \right)}}{\omega_i}, \quad \delta_{\eta \pm} = \frac{\omega_i \theta_{i\pm}}{2m_*}.	
\end{equation}
It can be seen from the expression (\ref{eq10}) that there are restrictions on the values of the quantum parameter $\varepsilon_{\eta\left(r\right)}$  and the outgoing angles of the positron (electron):
 
\begin{equation}\label{eq12}
\varepsilon_{\eta\left(r\right)}=r\varepsilon_\eta \ge 1, \quad \delta^2_{\eta \pm}	\le \delta^2_{\eta \max \left(r\right)}=\varepsilon_{\eta\left(r\right)}\left(\varepsilon_{\eta\left(r\right)}-1\right).	
\end{equation} 
It is important to note that the resonant energy of the positron and electron is determined by the corresponding  outgoing angle (ultrarelativistic parameter  ${\delta'}^2_{\eta \pm}$ (\ref{eq11})), as well as the quantum parameter $\varepsilon_{\eta\left(r\right)}$ (\ref{eq3}). Note that the resonant energy spectrum (\ref{eq10}) is essentially discrete, since each value of the number of absorbed laser photons corresponds to its resonant energy: $r \to E_{\eta \pm \left( r \right)}$ (\ref{eq10}). Note that the first relation in expression (\ref{eq12}), depending on the value of the quantum parameter  , can be represented as a condition for the number of the wave absorbed photons required for the resonant process:
\begin{equation}\label{eq13}
	r \ge r_{\min}=\lceil \varepsilon^{-1}_{\eta} \rceil, \ if \ \varepsilon_{\eta} < 1 \ \left(\omega_i<\omega_\eta\right),
\end{equation}
\begin{equation}\label{eq14}
	r \ge 1, \ if \ \varepsilon_{\eta} \ge 1 \ \left(\omega_i \ge \omega_\eta\right),
\end{equation}
 Figure~\ref{figure2} shows the resonant energy of a positron (for channel A) or an electron (for channel B) as a function of its square of the outgoing angle with a fixed number of absorbed photons of the wave (\ref{eq10}). 
 
 Solid lines correspond to the maximum resonant energy (the "+" sign before the square root in the ratio (\ref{eq10})). The dashed lines correspond to the minimum resonant energy (the "-" sign in front of the square root in the ratio (\ref{eq10})). From this figure it can be seen that with an increase in the number of absorbed photons of the wave, the maximum resonant energy, as well as the maximum outgoing angle, increase (for a fixed outgoing angle).
 
 We present the values of the characteristic energy of the process $\omega_\eta$ (\ref{eq4}) for various frequencies and intensities of the external electromagnetic wave. So, for the case of the flux of gamma quanta moving towards the direction of propagation of the electromagnetic wave $\left(\theta_i=\pi\right)$ , we obtain:
\begin{equation}\label{eq15}
\omega_\eta \approx \begin{cases}
	523.9\mbox{GeV},  \mbox{if} \ \omega=1  \mbox{eV}, \ \ \ \ \ I=1.861\cdot 10^{18}\mbox{Wcm}^{-2}; \\
	52.39\mbox{GeV},  \mbox{if} \ \omega=10 \mbox{eV}, \ \  \ I=1.861\cdot 10^{20}\mbox{Wcm}^{-2}; \\
    5.239\mbox {GeV}, \mbox{if} \ \omega=0.1\mbox{keV}, I=1.861\cdot 10^{22}\mbox{Wcm}^{-2}; \\
     1.31\mbox{GeV},  \ \ \mbox{if} \ \omega=1  \mbox{keV}, \ \ \  I=7.452\cdot 10^{24}\mbox{Wcm}^{-2}; \\
	 0.26\mbox{GeV}, \ \ \mbox{if} \ \omega=10 \mbox{keV}, \   I=1.675\cdot 10^{27}\mbox{Wcm}^{-2}. \\
		\end{cases}.
\end{equation} 
From this it can be seen that in the optical frequency domain $\left(\omega \sim 1 \mbox{eV}\right)$ , the minimum characteristic energy of the process has an order of magnitude $\sim 500 \mbox{GeV} \left(\eta \lesssim 1\right)$ . With an increase in the frequency of the external electromagnetic wave, the characteristic energy of the process decreases. The resonant process of photogeneration of pairs for optical laser frequencies in the region  (\ref{eq13}) was studied in detail in the article \cite{36}.

Here, within the framework of the relation (\ref{eq14}), we will study the case very high energies of the initial gamma quanta (\ref{eq5}). Because of this, for not very large outgoing angles, the resonant energies of the positron and electron (\ref{eq10}) will take the form:  
\begin{equation}\label{eq16}
	x_{\eta \pm \left( r \right)} \approx 1-\frac{\left(1+4\delta^2_{\eta \pm}\right)}{4\varepsilon_{\eta \left(r\right)}} \approx 1, \quad \left(\delta^2_{\eta \pm}\ll \varepsilon_{\eta \left(r\right)}, \ \varepsilon_{\eta \left(r\right)}\gg 1 \right).
\end{equation}
Here we have taken into account the maximum energy of the positron (electron). From here it can be seen that one of the possible values of the positron (electron) energies is close to the energy of the initial gamma quantum.  

\section{Maximum resonant cross section of the PGP in a strong field}

It is important to note that the resonant energy of the electron-positron pair is determined for channel A by the positron outgoing angle, and for channel B by the electron outgoing angle. In addition, channels A and B of the resonant FRP process do not interfere with each other. Because of this, the resonant differential cross section for channel A can be integrated at the electron outgoing angles, and for channel B at the positron outgoing angles.

In the article \cite{36}, a general relativistic expression was obtained for the resonant differential cross-section of the FRP process in the field of a strong electromagnetic wave with intensities up to $10^{27} \mbox{Wcm}^{-2}$. Integrations were carried out on the energies of the electron (positron), as well as the outgoing angles of the electron (for channel A) or positron (for channel B), on which the resonant energy of the positron (channel A) or electron (channel B) does not depend. Moreover, the integration at the outgoing angles was carried out in a special kinematic region, in which small relativistic corrections of the order $\left(m_*/\omega_i\right)^2 \ll1$ in the momentum transmitted to the nucleus are taken into account \cite{32,33,34,35,36,37,38,55}. It is this integration that leads to the appearance of a large order parameter $\left(\omega_i/m_*\right)^2 \gg1$  in the resonant differential cross-section \cite{32,33,34,35,36,37,38}. Also assume that the flux of initial gamma quanta is directed opposite to the direction of propagation of the electromagnetic wave. Taking this into account, the resonant differential cross section of the FRP with simultaneous registration of the energy and outgoing angles of the positron (sign "+" in equation (\ref{eq17}), channel A) or electron (sign "-" in equation (\ref{eq17}), channel B) can be represented as follows:   
\begin{equation}\label{eq17}
	R^{\max}_{\eta\pm\left(r\right)}=\frac{d\sigma^{\max}_{\eta\pm\left(r \right)}}{dx_{\eta\pm\left(r\right)}d\delta^2_{\eta \pm}}=\left(Z^2\alpha r^2_e\right)c_{\eta}\mathrm{H}_{\eta\pm\left(r \right)}.
\end{equation} 
Here $\alpha $ is the fine structure constant, $Z$ is the charge of the nucleus, $r_e$ is the classical radius of the electron, $\mathrm{H}_{\eta \pm \left( r \right)}$ are functions that determine the energy spectrum and angular distribution of a positron or electron:
\begin{equation}\label{eq18}
\mathrm{H}_{\eta \pm \left( r \right)}=\frac{x^3_{\eta\pm\left(r\right)}\left(1-x_{\eta\pm\left(r\right)}\right)^3}{\rho^2_{\eta\pm\left(r\right)}}P\left(u_{\eta\pm\left(r\right)},\varepsilon_{\eta \left(r\right)}\right),	
\end{equation}
\begin{equation}\label{eq19}
\rho_{\eta\pm\left(r\right)}=x^2_{\eta\pm\left(r\right)}\delta^2_{i \pm}+\frac{1}{4\left(1+\eta^2\right)}.
\end{equation}
and the magnitude of the $c_\eta$  coefficient is determined by the small transmitted momenta of the order $\left(m_*/\omega_i\right)^2 \ll1$, as well as the resonance width
\begin{equation}\label{eq20}
c_\eta=2\left[\frac{2\pi\omega_i}{\alpha m_* \mathrm{K}\left(\varepsilon_{\eta}\right)}\right]^2\gg1.
\end{equation}
Here the $\mathrm{K}\left(\varepsilon_{\eta}\right)$  function is determined by the resonance width (the full probability of the external field-stimulated Compton effect) and has the form \cite{23}:
\begin{eqnarray}\label{eq21}
	\mathrm{K} \left(\varepsilon_\eta \right)=\sum_{r=1}^\infty \mathrm K_r \left(\varepsilon_\eta \right),\
	\mathrm K_r \left(\varepsilon_\eta \right)=\int\limits_0^{\varepsilon_{\eta \left( r \right)}} \frac{du}{\left(1+u\right)^2} K \left(u, \varepsilon_{\eta \left( r \right)} \right).
\end{eqnarray}
\begin{eqnarray}\label{eq22}
	K\left(u, \varepsilon_{\eta \left( r \right)} \right)=-4J^2_r\left( \gamma_{\eta\left( r \right)} \right)+\eta^2 \left(2+\frac{u^2}{1+u} \right) \left( J^2_{r+1}+J^2_{r-1}-2J^2_r \right),
\end{eqnarray}
\begin{equation}\label{eq23}
	\gamma_{\eta \left( r \right)}=2r\frac{\eta }{\sqrt{1+\eta ^2}}\sqrt{\frac{u}{\varepsilon_{\eta \left( r \right)}}\left( 1-\frac{u}{\varepsilon_{\eta \left( r \right)}} \right)}.
\end{equation}
In expression (\ref{eq18}) the $P\left(u_{\eta\pm\left(r\right)},\varepsilon_{\eta \left( r \right)}\right)$ functions are determined by the probability (per unit of time) of the external field-stimulated Breit-Wheeler process \cite{23}:
\begin{eqnarray}\label{eq24}
P\left(u_{\eta\pm\left(r\right)},\varepsilon_{\eta \left( r \right)}\right)=J^2_r\left( \gamma_{\eta \pm \left(r\right)} \right)+\eta^2 \left(2u_{\eta\pm\left(r\right)}-1\right)\left[\left(\frac{r^2}{\gamma^2_{\eta \pm \left(r\right)}}-1\right)J^2_r+\frac{1}{4}\left(J_{r-1}-J_{r+1}\right)^2\right]. 
\end{eqnarray}
Here, the relativistically invariant parameter $u_{\eta\pm\left(r\right)}$  and the arguments of the Bessel functions $\gamma_{\eta \pm\left(r\right)}$  have the form:
\begin{equation}\label{eq25}
	u_{\eta \pm \left( r \right)}\approx \frac{1}{4x_{\eta \pm \left( r \right)}\left(1-x_{\eta \pm \left( r \right)}\right)}.
\end{equation}
\begin{eqnarray}\label{eq26}
	\gamma_{\eta \pm \left( r \right)}=2r\frac{\eta }{\sqrt{1+\eta ^2}}\sqrt{\frac{u_{\eta \pm \left( r \right)}}{\varepsilon_{\eta \left( r \right)}}\left( 1-\frac{u_{\eta \pm \left( r \right)}}{\varepsilon_{\eta \left( r \right)}} \right)}=4r\frac{\eta }{\sqrt{1+\eta ^2}}\frac{x_{\eta \pm \left( r \right)}\delta_{i \pm}}{\left(1+4x^2_{\eta \pm \left( r \right)}\delta^2_{i \pm}\right)}.
\end{eqnarray}
Note that the right part of the expression (\ref{eq26}) for the argument  of the Bessel functions  is obtained by taking into account the relations (\ref{eq10}) and (\ref{eq25}).

These resonant differential cross sections (\ref{eq17}) were studied in detail for the case when the energy of the initial gamma quanta did not exceed the characteristic energy of the process (\ref{eq13}). However, the most interesting case when the energy of the initial gamma quanta exceeds the characteristic energy of the process (\ref{eq14}) has not been studied. Here we consider the maximum resonant differential cross section (\ref{eq17}) for the case when the energy of the initial gamma quanta exceeds the characteristic energy of the process (\ref{eq14}). 

It is also of interest to consider the case (\ref{eq5}) when the energy of the initial gamma quanta significantly exceeds the characteristic energy of the process $\left(\omega_i \gg \omega_\eta \right)$ . In this case, the resonant energy of the positron or electron is close to the energy of the initial gamma quanta (see relation (\ref{eq16})). Given this, after simple transformations, the maximum resonant differential cross sections of the PGP  (\ref{eq17})-(\ref{eq25})  is significantly simplified and takes the form
\begin{equation}\label{eq27}
	R^{\max}_{\eta\pm\left(r\right)}=\frac{d\sigma^{\max}_{\eta\pm\left(r \right)}}{dx_{\eta\pm\left(r\right)}d\delta^2_{\eta \pm}}=\left(Z^2\alpha r^2_e\right)b_{\eta}\mathrm{\Phi}_{\eta\pm\left(r \right)}.
\end{equation}
Here the $\mathrm{\Phi}_{\eta\pm\left(r \right)}$ are functions determine the spectral-angular distribution of the resonant SB cross-section for channels A and B:
\begin{eqnarray} \label{eq28}
	\mathrm{\Phi}_{\eta\pm\left(r \right)}=\frac{\left(1+4\delta^2_{\eta \pm}\right)^3}{r^3}\left[\delta^2_{i \pm}+\frac{1}{4\left(1+\eta^2\right)}\right]^{-2}P\left(\delta^2_{\eta \pm},\varepsilon_{\eta \left( r \right)}\right)
\end{eqnarray}
and $b_\eta$ - the coefficient, which is determined by the parameters of the laser installation
\begin{equation}\label{eq29}
b_\eta=\frac{1}{2\varepsilon_{\eta}}\left(\frac{\pi\omega_\eta}{2 \alpha m_* \mathrm{K}\left(\varepsilon_{\eta}\right)}\right)^2.
\end{equation}
In expression (\ref{eq28}) the function (\ref{eq24}) takes the form:
\begin{eqnarray} \label{eq30}
P\left(\delta^2_{\eta \pm},\varepsilon_{\eta \left( r \right)}\right)=J^2_r\left( \gamma_{\eta \pm \left(r\right)} \right)+\eta^2 \left[\frac{2\varepsilon_{\eta \left( r \right)}}{\left(1+4\delta^2_{\eta \pm}\right)}-1\right]\left[\left(\frac{r^2}{\gamma^2_{\eta \pm \left(r\right)}}-1\right)J^2_r+\frac{1}{4}\left(J_{r-1}-J_{r+1}\right)^2\right], 	
\end{eqnarray}
\begin{eqnarray}\label{eq31}
	\gamma_{\eta \pm \left( r \right)}=4r\frac{\eta }{\sqrt{1+\eta ^2}}\frac{\delta_{\eta \pm}}{\left(1+4\delta^2_{\eta \pm}\right)}.
\end{eqnarray}
It is worth noting that the obtained expressions (\ref{eq17}) and (\ref{eq27}) are true for the case of the one initial gamma quantum. To derive the relations for the case of the gamma quantum flux one has to multiply the corresponding equations by the concentration $n_\gamma$.

\section{Main results}

Let the flux of initial gamma quanta propagate towards an external electromagnetic wave $\left(\theta_i=\pi\right)$. Let's choose the energy of the initial gamma quanta $\omega_i=60\mbox{GeV}$. Then, for characteristic energies $\omega_\eta$  (\ref{eq15}), the quantum parameter $\varepsilon_{\eta}$ (\ref{eq3}) takes the corresponding values: $\varepsilon_{\eta}=0.115$; $1.145$; $11.453$; $45.812$; $229.057$. Here, the first case corresponds to the optical frequencies of the laser $\left(\omega=1\mbox{eV}, \ I=1.863\cdot 10^{18} \mbox{Wcm}^{-2}\right)$ and meets the condition (\ref{eq13}) when $r\ge r_{\min}=9$. The remaining cases for the X-ray frequencies of the external wave meet the condition (\ref{eq14}) when $r\ge 1 \left(\omega_i\ge \omega_\eta\right)$. Moreover, the last three cases meet the condition $\varepsilon_{\eta}\gg1\left(\omega_i\gg \omega_\eta\right) $ (see Exp. (\ref{eq16})).

Note that when plotting the resonant differential cross-section (\ref{eq17}), (\ref{eq27}), the maximum energy of the positron (electron) was selected (the "+" sign before the square root in expression (\ref{eq10})). It is these positron (electron) energies that make the main contribution to the resonant differential cross section.

\begin{figure}[H]
		\centering
	\includegraphics[width=10cm]{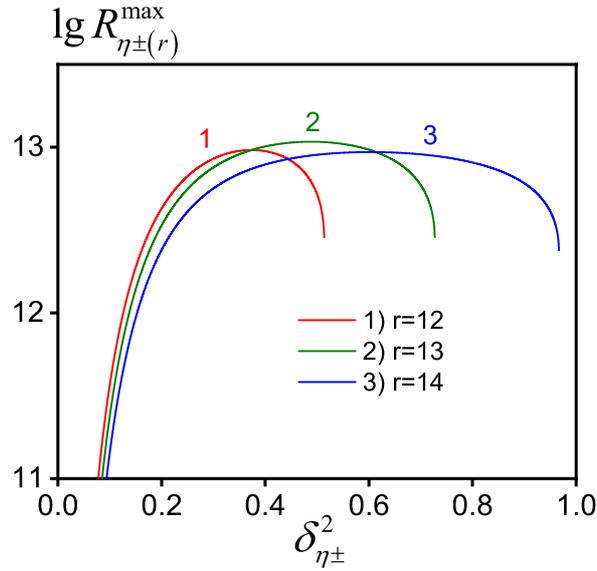}
	\caption{The dependence of the maximum resonant differential cross-section  (in  $Z^2\alpha r^2_e$ units) (\ref{eq17}) on the square of the outgoing angle of the positron (for channel A) or electron (for channel B) for a fixed number of absorbed photons of the wave when the value $r_{\min}=9$ (\ref{eq13}). The curves are constructed for the maximum energy of the positron (electron) (in the ratio (\ref{eq10}), the sign "+" is chosen before the square root).The energy of the initial gamma quanta is equal to $\omega_i=60\ \text{GeV}$. The direction of propagation, frequency and intensity of the wave are equal to $\theta_i=\pi$, $\omega=1\ \text{eV}$, $I=1.863\cdot 10^{18} \mbox{Wcm}^{-2}$.} 
	\label{figure3}
\end{figure}

Figure~\ref{figure3} shows the dependences of the maximum resonant differential cross section (\ref{eq17}) on the square of the positron (electron) outgoing angle for a fixed number of absorbed photons of the wave in the optical frequency range $\left(\omega=1\mbox{eV}, \ I=1.863\cdot 10^{18} \mbox{Wcm}^{-2}\right)$ under conditions (\ref{eq13}) when $r_{\min}=9$. It is important to note that in this case, the maximum value of the resonant differential cross section takes place at the number of absorbed photons $r=13$  and is the value $R^{\max}_{\eta \pm \left(13 \right)} \approx 10^{13} \left( Z^2\alpha r^2_e \right)$ . At the same time, the resonant energies of the positron and electron are approximately equal to half the energy of the initial gamma quanta (see Table~\ref{tab1}).

\begin{table}[h!]
	\caption{The values of the resonant energies of the electron-positron pair and the corresponding square outgoing angle of the positron (electron) for the maximum value of the maximum resonant cross section (see Figure~\ref{figure3}). The frequency and intensity of the laser wave are $\omega=1\ \text{eV}$ and $I=1.863\cdot 10^{18} \mbox{Wcm}^{-2}$. The energy of the initial gamma quanta is $\omega_i=60\ \text{GeV}$.\label{tab1}}
	\centering
	\setlength{\extrarowheight}{2mm}
	\begin{tabular}{|c|c|c|c|c|}
		\hline
		$r$ & $\delta^2_{\eta \pm}$ & $R^{\max}_{\eta \pm \left( r \right)}$ & $E_{\pm \left( r \right)}$ & $E_{\mp \left( r \right)}$ \\\hline
		& & $\left( Z^2\alpha r^2_e \right)$ & $\left(\mbox{GeV}\right)$ & $\left(\mbox{GeV}\right)$ \\ \hline
		12 & 0.374 & $9.623\times 10^{12}$ & 30.00000 &   30.00000\\\hline
		13 & 0.488 & $1.080\times 10^{13}$ & 30.01535 &   29.98465 \\\hline
		14 & 0.603 & $9.356\times 10^{12}$ & 30.00000 &   30.00000 \\\hline
	\end{tabular}
	\setlength{\extrarowheight}{0mm}
\end{table}

\begin{figure}[H]
	\centering
	\includegraphics[width=10cm]{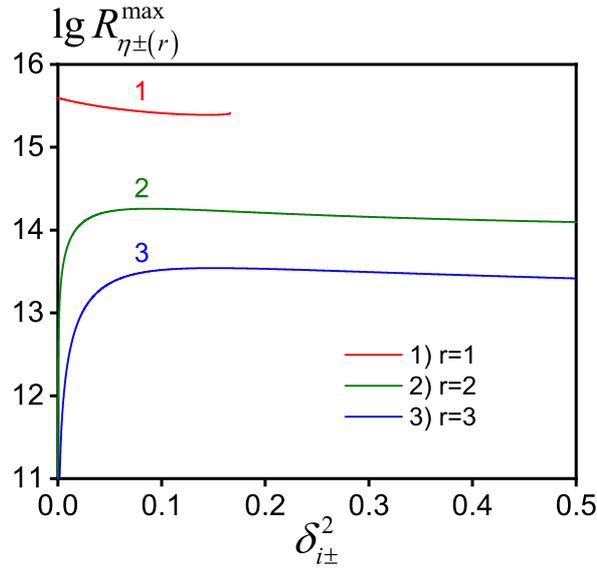}
	\caption{The dependence of the maximum resonant differential cross-section  (in  $Z^2\alpha r^2_e$ units) (\ref{eq17}) on the square of the outgoing angle of the positron (for channel A) or electron (for channel B) for a fixed number of absorbed photons of the wave in the conditions (\ref{eq14}). The curves are constructed for the maximum energy of the positron (electron) (in the ratio (\ref{eq10}), the sign "+" is chosen before the square root).The energy of the initial gamma quanta is equal to $\omega_i=60\ \text{GeV}$. The direction of propagation, frequency and intensity of the wave are equal to $\theta_i=\pi$, $\omega=10\ \text{eV}$, $I=1.863\cdot 10^{20} \mbox{Wcm}^{-2}$.} 
	\label{figure4}
\end{figure}

Figure~\ref{figure4} shows the dependences of the maximum resonant differential cross section (\ref{eq17}) on the square of the positron (electron) outgoing angle for a fixed number of absorbed photons of the wave $r=1,2,3$ in the X-ray frequency range $\left(\omega=10\mbox{eV}, \ I=1.863\cdot 10^{20} \mbox{Wcm}^{-2}\right)$ under conditions (\ref{eq14}). At the same time, the energy of the initial gamma quanta slightly exceeds the characteristic energy of the process $\left(\varepsilon_{\eta}=1.145\right)$. It is important to note that in this case, the maximum value of the resonant differential cross section occurs when one photon of the wave is absorbed at $\delta^2_{\eta \pm}=0$  and is of the order of magnitude $R^{\max}_{\eta \pm \left(1 \right)} \sim 10^{15} \left( Z^2\alpha r^2_e \right)$. With an increase in the number of absorbed photons of the wave, the peak of the maximum value of the resonant differential cross section shifts towards large outgoing angles of the positron (electron). At the same time, the resonant differential cross-section decreases quite quickly. Thus, the ratio of the resonant cross sections for three and one absorbed photons of the wave has the order of magnitude: $R^{\max}_{\eta \pm \left(3 \right)}/R^{\max}_{\eta \pm \left(1 \right)}\sim 10^{-2}$. Note also that the resonant energy of a positron (for channel A) or an electron (for channel B) increases from $40.685 \mbox{GeV}$  for $r=1$  to $52.727 \mbox{GeV}$  for $r=3$  with an increase in the number of absorbed photons of the wave (see Table~\ref{tab2}).

\begin{table}[h!]
	\caption{The values of the resonant energies of the electron-positron pair and the corresponding square outgoing angle of the positron (electron) for the maximum value of the maximum resonant cross section (see Figure~\ref{figure4}). The frequency and intensity of the laser wave are $\omega=10\ \text{eV}$ and $I=1.863\cdot 10^{20} \mbox{Wcm}^{-2}$. The energy of the initial gamma quanta is $\omega_i=60\ \text{GeV}$.\label{tab2}}
	\centering
	\setlength{\extrarowheight}{2mm}
	\begin{tabular}{|c|c|c|c|c|}
		\hline
		$r$ & $\delta^2_{\eta \pm}$ & $R^{\max}_{\eta \pm \left( r \right)}$ & $E_{\pm \left( r \right)}$ & $E_{\mp \left( r \right)}$ \\\hline
		& & $\left( Z^2\alpha r^2_e \right)$ & $\left(\mbox{GeV}\right)$ & $\left(\mbox{GeV}\right)$ \\ \hline
		1 & 0 & $3.937\times 10^{15}$ & 40.685 &   19.315\\\hline
		2 & 0.088 & $1.811\times 10^{14}$ & 50.238 &   9.762 \\\hline
		3 & 0.150 & $3.478\times 10^{13}$ & 52.727 &   7.273 \\\hline
	\end{tabular}
	\setlength{\extrarowheight}{0mm}
\end{table}

\begin{figure}[H]
	\centering
	\includegraphics[width=10cm]{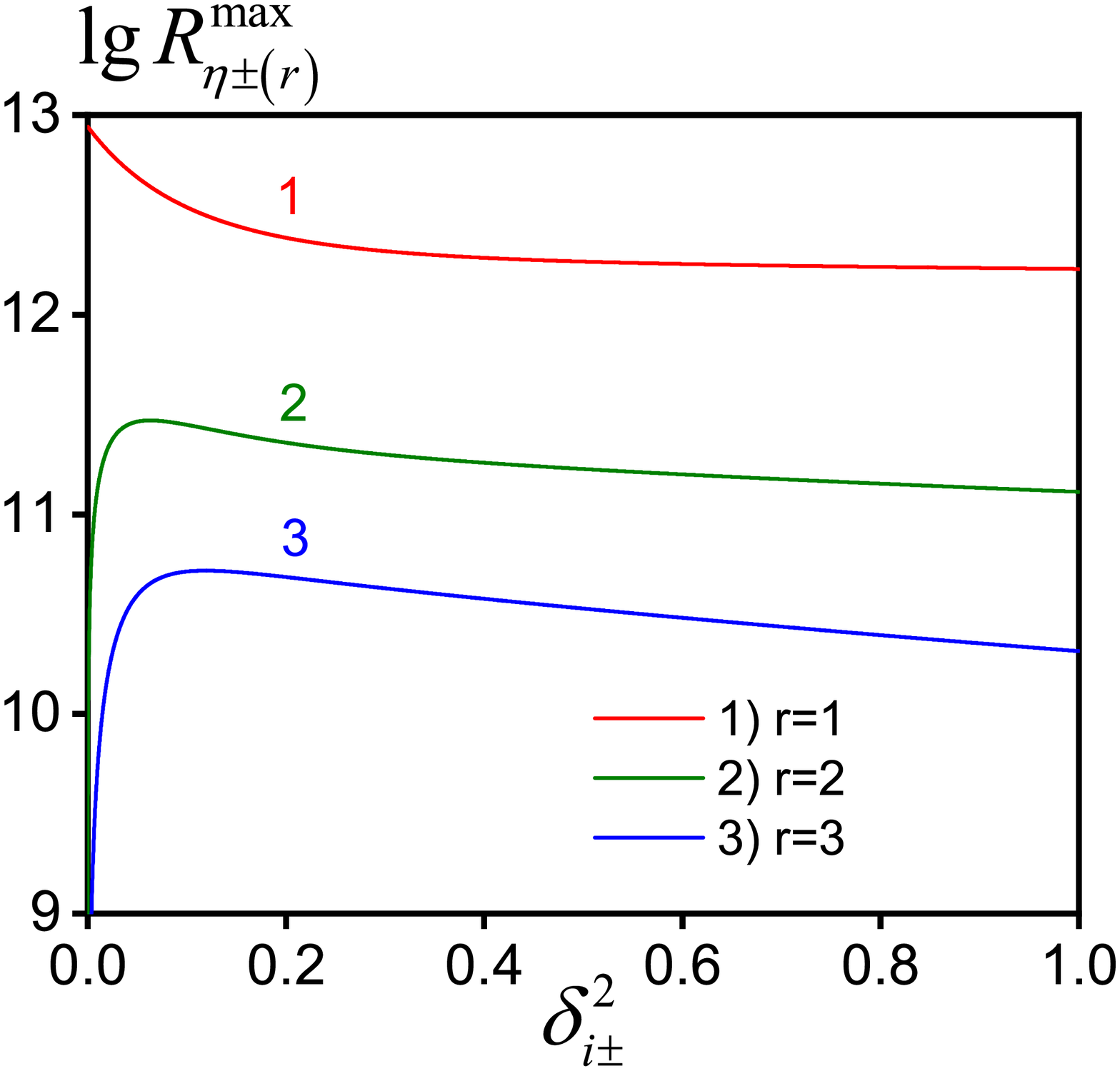}
	\caption{The dependence of the maximum resonant differential cross-section  (in  $Z^2\alpha r^2_e$ units) (\ref{eq17}), (\ref{eq27})  on the square of the outgoing angle of the positron (for channel A) or electron (for channel B) for a fixed number of absorbed photons of the wave in the conditions (\ref{eq14}), (\ref{eq16}). The curves are constructed for the maximum energy of the positron (electron) (in the ratio (\ref{eq10}), the sign "+" is chosen before the square root).The energy of the initial gamma quanta is equal to $\omega_i=60\ \text{GeV}$. The direction of propagation, frequency and intensity of the wave are equal to $\theta_i=\pi$, $\omega=0.1\ \text{keV}$, $I=1.863\cdot 10^{22} \mbox{Wcm}^{-2}$.} 
	\label{figure5}
\end{figure}

\begin{figure}[H]
	\centering
	\includegraphics[width=10cm]{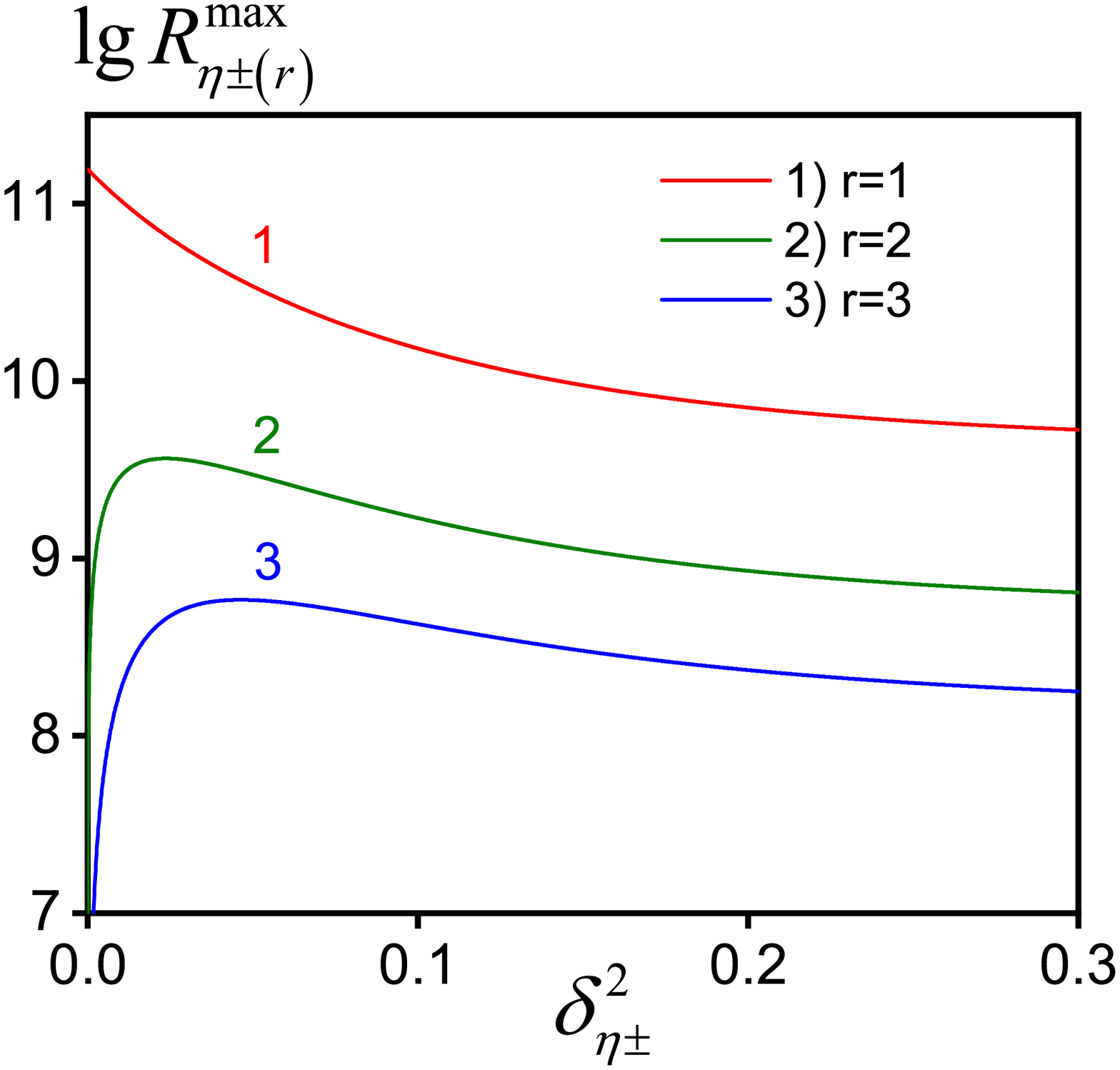}
	\caption{The dependence of the maximum resonant differential cross-section  (in  $Z^2\alpha r^2_e$ units) (\ref{eq17}), (\ref{eq27}) on the square of the outgoing angle of the positron (for channel A) or electron (for channel B) for a fixed number of absorbed photons of the wave in the conditions (\ref{eq14}), (\ref{eq16}). The curves are constructed for the maximum energy of the positron (electron) (in the ratio (\ref{eq10}), the sign "+" is chosen before the square root).The energy of the initial gamma quanta is equal to $\omega_i=60\ \text{GeV}$. The direction of propagation, frequency and intensity of the wave are equal to $\theta_i=\pi$, $\omega=1\ \text{keV}$, $I=7.452\cdot 10^{24} \mbox{Wcm}^{-2}$.} 
	\label{figure6}
\end{figure}

\begin{figure}[H]
	\centering
	\includegraphics[width=10cm]{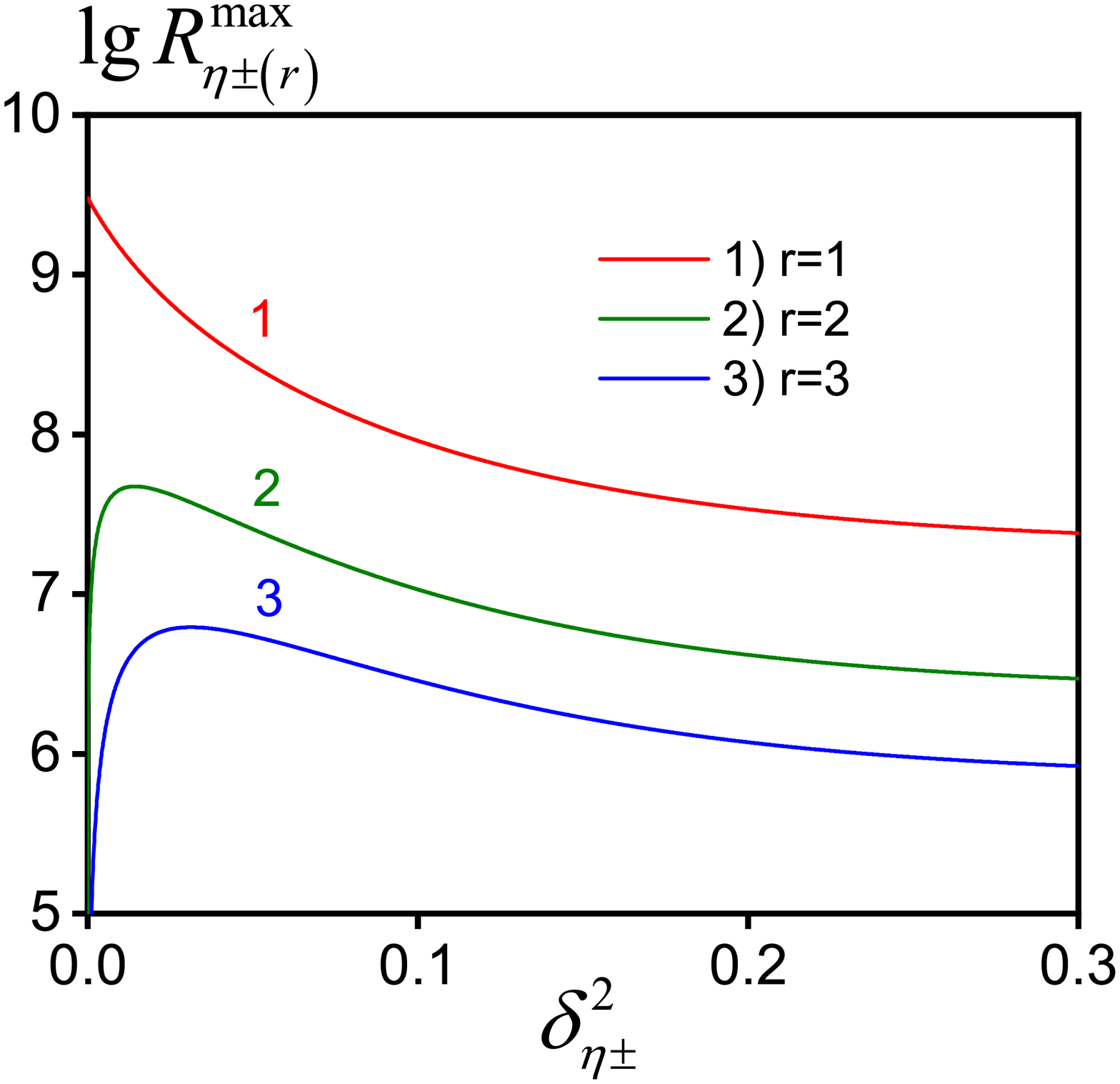}
	\caption{The dependence of the maximum resonant differential cross-section  (in  $Z^2\alpha r^2_e$ units) (\ref{eq17}), (\ref{eq27}) on the square of the outgoing angle of the positron (for channel A) or electron (for channel B) for a fixed number of absorbed photons of the wave in the conditions (\ref{eq14}), (\ref{eq16}). The curves are constructed for the maximum energy of the positron (electron) (in the ratio (\ref{eq10}), the sign "+" is chosen before the square root).The energy of the initial gamma quanta is equal to $\omega_i=60\ \text{GeV}$. The direction of propagation, frequency and intensity of the wave are equal to $\theta_i=\pi$, $\omega=10\ \text{keV}$, $I=1.676\cdot 10^{27} \mbox{Wcm}^{-2}$.} 
	\label{figure7}
\end{figure}

Figures~\ref{figure5}, \ref{figure6}, \ref{figure7} show the dependences of the maximum resonant differential cross-section (\ref{eq17}), (\ref{eq27}) on the square of the positron (electron) outgoing angle for a fixed number of absorbed photons of the wave $\left(r=1,2,3\right)$ for X-ray frequencies $\omega=0.1\ \text{keV}$, $1\ \text{keV}$, $10\ \text{keV}$ and corresponding wave intensities (see expression (\ref{eq15})). These graphs are constructed under conditions when the energy of the initial gamma quanta significantly exceeds the characteristic energy of the process: $\varepsilon_{\eta}\approx11.45$, $45.81$, $229.06$  (see the ratios (\ref{eq14}), (\ref{eq15}), (\ref{eq16})).

Tables~\ref{tab3}, \ref{tab4}, \ref{tab5} show the outgoing angles, the resonant energies of the positron (electron), as well as the values of the resonant differential cross sections corresponding to the maxima of the distributions in graphs~\ref{figure5}, \ref{figure6}, \ref{figure7}. From these figures and tables it can be seen that the maximum value of the resonant differential cross section is realized with one absorbed photon at $\delta^2_{\eta \pm}=0$ . With an increase in the number of absorbed photons, as well as the intensity of the wave, the value of the maximum resonant differential cross section decreases.

\begin{table}[h!]
	\caption{The values of the resonant energies of the electron-positron pair and the corresponding square outgoing angle of the positron (electron) for the maximum value of the maximum resonant cross section (see Figure~\ref{figure5}). The frequency and intensity of the laser wave are $\omega=0.1\ \text{keV}$ and $I=1.863\cdot 10^{22} \mbox{Wcm}^{-2}$. The energy of the initial gamma quanta is $\omega_i=60\ \text{GeV}$.\label{tab3}}
	\centering
	\setlength{\extrarowheight}{2mm}
	\begin{tabular}{|c|c|c|c|c|}
		\hline
		$r$ & $\delta^2_{\eta \pm}$ & $R^{\max}_{\eta \pm \left( r \right)}$ & $E_{\pm \left( r \right)}$ & $E_{\mp \left( r \right)}$ \\\hline
		& & $\left( Z^2\alpha r^2_e \right)$ & $\left(\mbox{GeV}\right)$ & $\left(\mbox{GeV}\right)$ \\ \hline
		1 & 0 & $8.754\times 10^{12}$ & 58.660 &   1.340\\\hline
		2 & 0.063 & $2.952\times 10^{11}$ & 59.173 &   0.827 \\\hline
		3 & 0.118 & $5.221\times 10^{10}$ & 59.353 &   0.647 \\\hline
	\end{tabular}
	\setlength{\extrarowheight}{0mm}
\end{table}

So, for the wave intensities $I=1.863\cdot 10^{22}$, $7.452\cdot 10^{24}$, $1.675\cdot 10^{27} \ \mbox{Wcm}^{-2}$, as well as for the number of absorbed photons $r=1$ and $r=2$  the value of the maximum resonant differential cross section, respectively, is equal to $R^{\max}_{\eta \pm \left( 1 \right)}\approx 8.75\cdot 10^{12}$, $1.56\cdot 10^{11}$, $3.03\cdot 10^9 \left( Z^2\alpha r^2_e \right)$ and $R^{\max}_{\eta \pm \left( 2 \right)}\approx 2.95\cdot 10^{11}$, $3.66\cdot 10^9$, $4.73\cdot 10^7 \left( Z^2\alpha r^2_e \right)$ (see Tables~\ref{tab3}, \ref{tab4}, \ref{tab5}). It is very important to note that with an increase in parameter $\varepsilon_{\eta}$ (\ref{eq3}), the value of the resonant energy of the positron (for channel A) or electron (for channel B) all it tends closer to the energy of the initial gamma quanta $E_i=60\mbox{GeV}$ (\ref{eq16}). So, for the wave intensities $I=1.863\cdot 10^{22}$, $7.452\cdot 10^{24}$, $1.675\cdot 10^{27} \ \mbox{Wcm}^{-2}$, as well as for the number of absorbed photons $r=1$ and $r=2$ the magnitude of the resonant energy of the positron (electron) at the maximum of the distribution of the resonant differential cross section (see Figures~\ref{figure5}, \ref{figure6}, \ref{figure7} ), respectively, is equal to $E_{\eta \pm \left( 1 \right)}\approx 58.660$, $59.671$, $59.934 \mbox{GeV}$  and  $E_{\eta \pm \left( 2 \right)}\approx 59.173$, $59.820$, $59.965 \mbox{GeV}$ (see Tables~\ref{tab3}, \ref{tab4}, \ref{tab5}).

Thus, under conditions (\ref{eq16}), when the energy of gamma quanta significantly exceeds the characteristic energy of the process, it is very likely to obtain narrowly directed streams of positrons (electrons) with energies close to the energy of the initial gamma quanta.

\begin{table}[h!]
	\caption{The values of the resonant energies of the electron-positron pair and the corresponding square outgoing angle of the positron (electron) for the maximum value of the maximum resonant cross section (see Figure~\ref{figure6}). The frequency and intensity of the laser wave are $\omega=1\ \text{keV}$ and $I=7.452\cdot 10^{24} \mbox{Wcm}^{-2}$. The energy of the initial gamma quanta is $\omega_i=60\ \text{GeV}$.\label{tab4}}
	\centering
	\setlength{\extrarowheight}{2mm}
	\begin{tabular}{|c|c|c|c|c|}
		\hline
		$r$ & $\delta^2_{\eta \pm}$ & $R^{\max}_{\eta \pm \left( r \right)}$ & $E_{\pm \left( r \right)}$ & $E_{\mp \left( r \right)}$ \\\hline
		& & $\left( Z^2\alpha r^2_e \right)$ & $\left(\mbox{GeV}\right)$ & $\left(\mbox{GeV}\right)$ \\ \hline
		1 & 0 & $1.564\times 10^{11}$ & 59.671 &   0.329\\\hline
		2 & 0.023 & $3.661\times 10^9$ & 59.820 &  0.180 \\\hline
		3 & 0.046 & $5.844\times 10^8$ & 59.870 &   0.130 \\\hline
	\end{tabular}
	\setlength{\extrarowheight}{0mm}
\end{table}

\begin{table}[h!]
	\caption{The values of the resonant energies of the electron-positron pair and the corresponding square outgoing angle of the positron (electron) for the maximum value of the maximum resonant cross section (see Figure~\ref{figure6}). The frequency and intensity of the laser wave are $\omega=10\ \text{keV}$ and $I=1.676\cdot 10^{27} \mbox{Wcm}^{-2}$. The energy of the initial gamma quanta is $\omega_i=60\ \text{GeV}$.\label{tab5}}
	\centering
	\setlength{\extrarowheight}{2mm}
	\begin{tabular}{|c|c|c|c|c|}
		\hline
		$r$ & $\delta^2_{\eta \pm}$ & $R^{\max}_{\eta \pm \left( r \right)}$ & $E_{\pm \left( r \right)}$ & $E_{\mp \left( r \right)}$ \\\hline
		& & $\left( Z^2\alpha r^2_e \right)$ & $\left(\mbox{GeV}\right)$ & $\left(\mbox{GeV}\right)$ \\ \hline
		1 & 0 & $3.029\times 10^9$ & 59.934 &   0.066\\\hline
		2 & 0.014 & $4.730\times 10^7$ & 59.965 &  0.035 \\\hline
		3 & 0.031 & $6.209\times 10^6$ & 59.975 &   0.025 \\\hline
	\end{tabular}
	\setlength{\extrarowheight}{0mm}
\end{table}

\section{Conclusions}

The study of the resonant process of generation of ultrarelativistic electron-positron pairs by high-energy gamma quanta in the field of the nucleus and a strong electromagnetic wave showed: 

\begin{itemize}
	\item	The resonant energy of an electron-positron pair is determined by two parameters: the outgoing angle of a positron (for channel A) or an electron (for channel B), as well as a quantum parameter $\varepsilon_{\eta \left( r \right)}=r\varepsilon_{\eta} \ge 1$ equal to the number of absorbed photons of the wave multiplied by the quantum parameter $\varepsilon_{\eta}$. This parameter is equal to the ratio of the energy of the initial gamma quanta $\left(\omega_i\right)$ to the characteristic energy of the process $\left(\omega_\eta\right)$. The characteristic energy is determined by the parameters of the laser installation: frequency, intensity, as well as the direction of propagation of the electromagnetic wave relative to the momentum of the initial gamma quanta (\ref{eq3}), (\ref{eq4}).
	\item	The magnitude of the quantum parameter significantly affects the probability and energy of the electron-positron pair. So, if the energy of the initial gamma quanta is less than the characteristic energy $\left(\omega_i < \omega_\eta\right)$, then $\varepsilon_{\eta}<1$. In this case, there is a minimum number of photons $\left(r \ge r_{\min}=\lceil \varepsilon^{-1}_{\eta} \rceil\right)$, starting from which photons are absorbed by the wave in the external field-stimulated Breit-Wheeler process. In strong fields $r_{\min}\gg1$, the resonance process also proceeds with the absorption of a very large number of photons of the wave. 
	\item	If the energy of the initial gamma quanta is equal to or exceeds the characteristic energy of the process $\left(\omega_i \ge \omega_\eta\right)$ , then the quantum parameter $\varepsilon_{\eta}\ge1$. In this case, the resonant process takes place for the number of absorbed photons of the wave $r\ge1$. Note that the probability of processes with the absorption of a small number of photons of the wave $\left(r\sim1\right)$  significantly exceeds the corresponding probability with the absorption of a large number of photons of the wave $\left(r\gg1\right)$.
	\item	If the energy of the initial gamma quanta significantly exceeds the characteristic energy of the process $\varepsilon_{\eta}\gg1$ , then the resonant energy of the positron (for channel A) or electron (for channel B) will be close to the energy of gamma quanta (see relation (\ref{eq16})). Under these conditions, narrow streams of high-energy positrons (electrons) are generated with a very high probability. So, for the number of absorbed photons $r=1$ and $r=2$  the value of the maximum resonant differential cross section, respectively, is equal to $R^{\max}_{\eta \pm \left( 1 \right)}\approx 8.75\cdot 10^{12}$, $1.56\cdot 10^{11}$, $3.03\cdot 10^9 \left( Z^2\alpha r^2_e \right)$ and $R^{\max}_{\eta \pm \left( 2 \right)}\approx 2.95\cdot 10^{11}$, $3.66\cdot 10^9$, $4.73\cdot 10^7 \left( Z^2\alpha r^2_e \right)$ (see Tables~\ref{tab3}, \ref{tab4}, \ref{tab5}).
\end{itemize}


\begin{thebibliography}{99}
   
\bibitem{1}C. N. Danson, C. Haefner, J. Bromage, T. Butcher, J.-C. F. Chanteloup, E. A. Chowdhury, A. Galvanauskas, L. A. Gizzi, J. Hein, D. I. Hillier $et \: al.$, High Power Laser Science and Engineering 7, e54 (2019).
\bibitem{2} J. W. Yoon, Y. G. Kim, I. W. Choi, J. H. Sung, H. W. Lee, S. K. Lee, and C. H. Nam, Optica 8, 630 (2021).
\bibitem{3} I. C. E. Turcu, B. Shen, D. Neely, G. Sarri, K. A. Tanaka, P. McKenna, S. P. D. Mangles, T.-P. Yu, W. Luo, X.-L. Zhu, and  et al, High Power Laser Science and Engineering 7, e10 (2019).
\bibitem{4} K. A. Tanaka, K. M. Spohr, D. L. Balabanski, S. Balascuta, L. Capponi, M. O. Cernaianu, M. Cuciuc, A. Cucoanes, I. Dancus, A. Dhal, B. Diaconescu, D. Doria, P. Ghenuche, D. G. Ghita, S. Kisyov, V. Nastasa, J. F. Ong, F. Rotaru, D. Sangwan, P.-A. Söderström, D. Stutman, G. Suliman, O. Tesileanu, L. Tudor, N. Tsoneva, C. A. Ur, D. Ursescu, and N. V. Zamfir, Matter and Radiation at Extremes 5, 024402 (2020).
\bibitem{5} S. Weber, S. Bechet, S. Borneis, L. Brabec, M. Bučka, E. Chacon-Golcher, M. Ciappina, M. DeMarco, A. Fajstavr, K. Falk, E.-R. Garcia, J. Grosz, Y.-J. Gu, J.-C. Hernandez, M. Holec, P. Janečka, M. Jantač, M. Jirka, H. Kadlecova, D. Khikhlukha, O. Klimo, G. Korn, D. Kramer, D. Kumar, T. Lastovička, P. Lutoslawski, L. Morejon, V. Olšovcová, M. Rajdl, O. Renner, B. Rus, S. Singh, M. Šmid, M. Sokol, R. Versaci, R. Vrána, M. Vranic, J. Vyskočil, A. Wolf, and Q. Yu, Matter and Radiation at Extremes 2, 149 (2017).
\bibitem{6} D. N. Papadopoulos, J. P. Zou, C. L. Blanc, G. Chériaux, P. Georges, F. Druon, G. Mennerat, P. Ramirez, L. Martin, A. Fréneaux, and  et al, High Power Laser Science and Engineering 4, e34 (2016).    
\bibitem{7} J. Bromage, S.-W. Bahk, I. A. Begishev, C. Dorrer, M. J. Guardalben, B. N. Hoffman, J. B. Oliver, R. G. Roides, E. M. Schiesser, M. J. S. III $et \: al.$, High Power Laser Science and Engineering 7, e4 (2019).
\bibitem{8} J. Rossbach, J. R. Schneider, and W. Wurth, Physics Reports 808, 1 (2019).
\bibitem{9} A. Gonoskov, A. Bashinov, S. Bastrakov, E. Efimenko, A. Ilderton, A. Kim, M. Marklund, I. Meyerov, A. Muraviev, and A. Sergeev, Phys. Rev. X 7, 041003 (2017).
\bibitem{10} J. Magnusson, A. Gonoskov, M. Marklund, T. Z. Esirkepov, J. K. Koga, K. Kondo, M. Kando, S. V. Bulanov, G. Korn, and S. S. Bulanov, Phys. Rev. Lett. 122, 254801 (2019).
\bibitem{11} X.-L. Zhu, T.-P. Yu, M. Chen, S.-M. Weng, and Z.-M. Sheng, New Journal of Physics 20, 83013 (2018).
\bibitem{12} P. Musumeci, C. Boffo, S. S. Bulanov, I. Chaikovska, A. F. Golfe, S. Gessner, J. Grames, R. Hessami, Y. Ivanyushenkov, A. Lankford, G. Loisch, G. Moortgat-Pick, S. Nagaitsev, S. Riemann, P. Sievers, C. Tenholt, and K. Yokoya, Positron Sources for Future High Energy Physics Colliders, No. arXiv:2204.13245, arXiv, 2022.
\bibitem{13} A. Fedotov, A. Ilderton, F. Karbstein, B. King, D. Seipt, H. Taya, and G. Torgrimsson, Advances in QED with Intense Background Fields, No. arXiv:2203.00019, arXiv, 2022.
\bibitem{14} A. Di Piazza, C. Müller, K. Z. Hatsagortsyan, and C. H. Keitel, Rev. Mod. Phys. 84, 1177 (2012).
\bibitem{15} A. Gonoskov, T. G. Blackburn, M. Marklund, and S. S. Bulanov, Charged Particle Motion and Radiation in Strong Electromagnetic Fields, No. arXiv:2107.02161, arXiv, 2022.
\bibitem{16} T. G. Blackburn, Reviews of Modern Plasma Physics 4, 5 (2020).
\bibitem{17} Y. I. Salamin, S. X. Hu, K. Z. Hatsagortsyan, and C. H. Keitel, Physics Reports 427, 41 (2006).
\bibitem{18} F. Ehlotzky, K. Krajewska, and J. Z. Kamiński, Reports on Progress in Physics 72, 46401 (2009).
\bibitem{19} F. Cajiao Vélez, J. Z. Kamiński, and K. Krajewska, Atoms 7, 34 (2019).
\bibitem{20} F. V. Bunkin and M. V. Fedorov, Sov. Phys. JETP 22, 844 (1966).
\bibitem{21} V. I. Ritus, J. Sov. Laser Res.; (United States) 6, (1985).
\bibitem{22} P. Zhang, S. S. Bulanov, D. Seipt, A. V. Arefiev, and A. G. R. Thomas, Physics of Plasmas 27, 050601 (2020).
\bibitem{23} V. I. Ritus and A. I. Nikishov, Quantum Electrodynamics Phenomena in the Intense Field, Vol. 111 (Nauka, Moscow, 1979).
\bibitem{24} V. P. Oleinik, Journal of Experimental and Theoretical Physics 25, 697 (1967).
\bibitem{25} V. P. Oleinik, Journal of Experimental and Theoretical Physics 26, 1132 (1968).
\bibitem{26} J. Bos, W. Brock, H. Mitter, and T. Schott, Journal of Physics A: Mathematical and General 12, 715 (1979).
\bibitem{27} A. V. Borisov, V. Ch. Zhukovskii, and P. A. Éminov, Soviet Physics Journal 23, 184 (1980).
\bibitem{28} A. V. Borisov, V. Ch. Zhukovskii, A. K. Nasirov, and P. A. Éminov, Soviet Physics Journal 24, 107 (1981).
\bibitem{29} S. P. Roshchupkin, Laser Physics 6, 837 (1996).
\bibitem{30} S. P. Roshchupkin, A. A. Lebed’, E. A. Padusenko, and A. I. Voroshilo, Laser Physics 22, 1113 (2012).
\bibitem{31} K. Krajewska, Laser Phys. 21, 1275 (2011).
\bibitem{32} N. R. Larin, V. V. Dubov, and S. P. Roshchupkin, Phys. Rev. A 100, 52502 (2019).
\bibitem{33} N. R. Larin, V. V. Dubov, and S. P. Roshchupkin, Modern Physics Letters A 35, 2040025 (2020).
\bibitem{34} S. P. Roshchupkin, N. R. Larin, and V. V. Dubov, Laser Physics 31, 45301 (2021).
\bibitem{35} N. R. Larin, S. P. Roshchupkin, and V. V. Dubov, Universe 6, (2020).
\bibitem{36} S. P. Roshchupkin, N. R. Larin, and V. V. Dubov, Phys. Rev. D 104, 116011 (2021).
\bibitem{37} S. P. Roshchupkin, A. V. Dubov, V. V. Dubov, and S. S. Starodub, New Journal of Physics 24, 13020 (2022).
\bibitem{38} S. Roshchupkin, A. Dubov, and S. Starodub, Universe 8, 218 (2022).
\bibitem{39} A. Hartin, Second Order QED Processes in an Intense Electromagnetic Field, No. arXiv:1701.02906, arXiv, 2017.
\bibitem{40} R. Ruffini, G. Vereshchagin, and S.-S. Xue, Physics Reports 487, 1 (2010).
\bibitem{41} E. Lötstedt, U. D. Jentschura, and C. H. Keitel, New Journal of Physics 11, 13054 (2009).
\bibitem{42} A. Di Piazza, E. Lötstedt, A. I. Milstein, and C. H. Keitel, Phys. Rev. A 81, 062122 (2010).
\bibitem{43} S. Augustin and C. Müller, Journal of Physics: Conference Series 497, 12020 (2014).
\bibitem{44} C. Müller, A. B. Voitkiv, and N. Grün, Phys. Rev. A 67, 063407 (2003).
\bibitem{45} K. Krajewska and J. Z. Kamiński, Phys. Rev. A 82, 013420 (2010).
\bibitem{46} B. Hafizi, D. F. Gordon, and D. Kaganovich, Phys. Rev. Lett. 122, 233201 (2019).
\bibitem{47} A. A. Lebed’, Laser Physics Letters 13, 45401 (2016).
\bibitem{48} A. Di Piazza and T. Pătuleanu, Phys. Rev. D 104, 076003 (2021).
\bibitem{49} F. Mackenroth and A. Di Piazza, Phys. Rev. Lett. 110, 070402 (2013).
\bibitem{50} D. Seipt and B. Kämpfer, Phys. Rev. D 85, 101701 (2012).
\bibitem{51} A. Ilderton, Phys. Rev. Lett. 106, 020404 (2011).
\bibitem{52} A. Hartin, International Journal of Modern Physics A 33, 1830011 (2018).
\bibitem{53} A. A. Mironov, S. Meuren, and A. M. Fedotov, Phys. Rev. D 102, 053005 (2020).
\bibitem{54} D. Volkov, Z. Phys 94, 250 (1935).
\bibitem{55} V. B. Berestetskii, E. M. Lifshitz, and L. P. Pitaevskii, Quantum Electrodynamics: Volume 4 (Butterworth-Heinemann, 1982).
	
\end{thebibliography}
\end{document}